# Remarkably high value of Capacitance in $BiFeO_3$ Nanorod

**Nabanita Dutta, S.K. Bandyopadhyay*, Subhasis Rana, Pintu Sen and A.K. Himanshu**

**Variable Energy Cyclotron Centre,**

**1/AF, Bidhan Nagar, Kolkata-700 064, India.**

***corresponding author. E-mail: skband@vecc.gov.in**

**Abstract:**

A remarkably high value of specific capacitance of 450 F/g has been observed through electrochemical measurements in the electrode made of multiferroic Bismuth Ferrite (BFO) in the form of nanorods protruding out. These BFO nanorods were developed on porous Anodised Alumina (AAO) templates using wet chemical technique. Diameters of nanorods were in the range of 20-100 nm. The high capacitance is attributed to the nanostructure. The active surface charge has been evaluated electrochemically by cyclic voltammetry (CV) at different scanning rates and charge-discharge studies. The specific capacitances were constant after several cycles of charge-discharge leading to their useful application in devices. The mechanism of accumulation of charge on the electrode surface has been studied.

## Introduction:

Electrochemical capacitors (EC), popularly known as supercapacitors provide high power and long cycle life, essential for energy storage devices. They are categorized as electrochemical double layer capacitors (EDLC) and pseudocapacitors. In EDLC, capacitance originates in the charge separation at the electrode-electrolyte interface, whereas pseudocapacitance arises from fast, reversible faradaic redox reactions taking place on or near the surface of the electrode [1]. Electrochemical performances of a material as electrode can be assayed by cyclic voltammetry and galvanostatic charge-discharge studies of specific capacitance. An electrode is judged by its capacitance value and the number of charge-discharge cycles it withstands, maintaining the constancy of capacitance. This brings forth the search for a wide variety of materials. In general, transition metal oxides like $RuO_2$, $MnO_2$ etc. show high specific capacitance with their redox behavior.

In this context, it is quite interesting to study the capacitance of multifunctional materials like the multiferroic Bismuth ferrite (BFO) showing coexistence of ferroelectricity as well as antiferromagnetism with wide applications [2]. Lokhande et al. [3] observed specific capacitance value of 81F/gm in BFO films. Attempts have been made to obtain BFO in 1-Dimensional nanostructure forms like nanowire, nanorod, etc. [4,5]. The nanostructured forms can be expected to offer better efficiency owing to large surface area giving rise to a high value of specific capacitance, for example in $\alpha$-$MnMoO_4$ nanorods [6]. This prompted us to study the capacitance of BFO nanostructure with its redox behaviour.

In this letter, we report the development of BFO nanorod by the wet chemical template assisted method and its capacitance studies by electrochemical means. We have evaluated specific capacitance both through cyclic voltammetry at different scanning rates and charge-discharge studies galvanostatically employing different currents. To our knowledge, there has not been any study of BFO nanorod in this respect so far.

**Experimental details:**

AAO templates of 60 μm thickness and 13 mm diameter with pore size distribution ranges from 20-100 nm were employed. 0.1M solutions of $Bi(NO_3)_3$ and $Fe(NO_3)_3$ were prepared with stoichiometric amounts of the nitrates using methoxymethanol as solvent with pH adjusted to 2-3. The filling of nanopores was achieved by the directional flow of the ions in the template adopting the controlled vacuum technique. The templates with pores containing solution were sintered for 3 hours at $750^0$C to get the required phase without unwanted grain growth. We attempted controlled etching process with 1M NaOH and the bundle of nanowires and nanorods emerged. Weights of BFO nanorods measured using a very sensitive microbalance (resolution of 1μgm) by subtracting the weights of respective AAO were in the range 80-100μg.

The nanowires and nanorods were examined by FEI Scanning Electron Microscope (SEM) with a resolution of 6nm aided by Energy Dispersive X-Ray (EDX) for compositional analysis. Transmission electron microscopy (TEM) was done by high resolution TEM (Model: FEI T20 with applied voltage of 200KV). Selective Area Electron Diffraction (SAED) was also undertaken to ascertain crystal structure of the nanorods.

Cyclic Voltammetry (CV) and galvanostatic charge-discharge were studied with AUTOLAB-30 potentiostat/galvanostat for both BFO on AAO and AAO blank templates being used as electrodes. The half etched templates used in SEM were employed for this purpose with the protrusion length of 1μm (as visible by SEM) over the template of thickness 60μm. Electrodes were prepared by connecting Cu lids with AAO/BFO templates through conducting silver paste. All the electrochemical experiments (i.e

CV, Charge-discharge) were performed with two-electrode system having identical electrodes with respect to shape, size and made of same active electrode materials (i.e. Type-I symmetric supercapacitor) using an electrolyte containing 1M $Na_2SO_4$ in water. A platinum electrode and a saturated Ag/AgCl electrode were used as counter and reference electrodes respectively. All the CVs were measured between -0.6 to +0.6 V (i.e. operating window of 1.2V) with respect to reference electrode at different scan rates (5mV/s to 50mV/s). Constant currents ranging from 15 to 30 μAmp have been employed for charging/discharging the cell in the voltage range from -0.6 to +0.6 V.

In a symmetrical system where the active material weight is the same for the two electrodes,

$$C_s = \frac{2C}{m} \tag{1}$$

where m is the active mass of the single electrode, $C$ is the discharge capacitance and $C_s$ is the specific capacitance of the electrode [7]. The charge accumulated on BFO was assayed by subtracting the charge on blank AAO from that on BFO/AAO electrode. We have applied several cycles of CV as well as charge- discharge to study the stability of the system with cycling.

**Results and Discussions:**

There are two kinds of 1D nanostructure in the form of nanowires as well as nanorods as apparent from figs. 1a and 1b similar to earlier groups [4,5]. Fig. 1a shows the bundles of nanowires and nanorods which have come out after etching the template with NaOH. Basically they are nanorods but we observed them in the form of a bundle after etching. There is a distribution in diameters of nanorods as revealed in fig. 1a. Cross sectional view is showing development of nanorods in Fig. 1b- a representative case. It demonstrates the structures of several nanorods (around 20 in a region of 5 μm x 5 μm) with high aspect ratio protruding out of the pores after partial etching. The compositional analysis was performed by EDX analysis at different nanorods and they reflected Bi:Fe atomic ratios of slightly more than 1:1 reflecting a little more Bi content with a fluctuation within 3  per cent. Fig. 1c shows the TEM picture of nanorods of high density with intact structure. The inset indicates a clear SAED pattern with prominent rings signifying the development of polycrystalline BFO.

Typical cyclic voltammograms (CV) of different samples at a scan rate of 50 and 10mV/s between -0.6 and 0.6V in aqueous solution of 1M $Na_2SO_4$ are shown in Fig. 2a. Cyclic voltammograms of different samples are quite symmetrical with a mirror image of the current response from voltage, indicating ideal pseudocapacitative behavior and excellent reversibility  in charging and discharging at a constant rate over the voltage range of –0.6 to 0.6V [8]. Voltammetric charges (q*) at different potential scan rate *v* (mV/s) were obtained by integration of the voltammetric curves followed by division with the geometric surface area of the samples without correction for background capacitative current. The charge accumulation on BFO/AAO and AAO individually assayed by CV established that the contribution from AAO was significantly less than BFO/AAO for same weight. Typical values were $2.270 \times 10^{-3}$ Coulombs for BFO/AAO and $6.45 \times 10^{-4}$ Coulombs for AAO at the scanning rate of 20mv/sec.

The applicability of the supercapacitor can also be evaluated by means of the galvanostatic charge–discharge studies. Charge–discharge profiles of different samples are shown in Fig. 2b. They exhibit a pseudocapacitative characteristic [9]. In general, the specific capacitance decreases gradually with increasing discharge current density due to increasing IR drops leading to higher d*v*/dt. We have observed the charge-discharge cycles at different currents and over quite substantial number of times – more than 100 and the times for charging and discharging are constant over long cycles for same current.

In cyclic voltammetry, the accumulation of charge on the electrode dominates on the surface. Total surface charge ($q^*_{total}$) related to the electrochemical active surface area of the electrode can be divided into two parts– $q^*_{out}$ and $q^*_{in}$.

$$q^*_{total} = q^*_{out} + q^*_{in} \qquad (2)$$

$q^*_{out}$ is the contribution of charge from the outer region of the electrodes directly exposed to the electrolyte and $q^*_{in}$ correlates with that from the inner part of the electrodes hidden in pores, grain boundary, etc. and reflects the regions of difficult accessibility for the ionic species assisting the surface redox reaction, essential for enhancing pseudocapacitance. Voltammetric charge q* depends on the potential scan rate $v$ [10]. The dependence is given by the relation:

$$q^*(v) = q^*_{\infty} + c\, v^{-1/2} \qquad (3)$$

with $q^*_{\infty}$ corresponding to $q^*_{out}$ in eqn. (2) and c is a constant of proportionality. At faster scan, the diffusion of ions is limited only to the more accessible sites, i.e. the outer surface of the electrode. Therefore, extrapolation of q* to scan rate $v = \propto$ (i.e. $v^{-1/2} = 0$) from the linear portion of the q* versus $v^{-1/2}$ plot can provide the accumulation of outer charge $q^*_{out}$ (from eqn. 3) related to the more easily accessible sites. On the other hand, q* can be extrapolated the other way round to $v = 0$ to extract $q_{tot}$. Since q* varies as $v^{-1/2}$, 1/q* is expected to behave linearly as $v^{1/2}$. Thus, the extrapolation of q* to the scan rate $v = 0$ in the plot 1/q* versus $v^{1/2}$ gives the total charge $q^*_{total}$, which is related to both inner and outer active sites of the electrode [10]. One can thus easily calculate the charge related to the inner sites (i.e. less accessible sites) $q^*_{in}$ of eqn. (2).

$$1/q^*(v) = (1/q_0^*) + c^* v^{1/2} \qquad (4)$$

The outer charge $q^*_{out}$ only will be relevant for extracting the specific capacitance of BFO nanorods. Fig. 3a shows q* versus $v^{-1/2}$ plot with linear fit. Extrapolation of q* to $v^{-1/2}=0$ gives the intercept 1.5 x $10^{-4}$ coulombs as $q_{out}$. Total charge obtained by extrapolating the plot of 1/q* vs. $v^{1/2}$ to $v^{1/2}$ =0 in Fig. 3b is 3.673 x $10^{-3}$ coulombs. Thus, $q^{*}_{in}$ = 3.523x$10^{-3}$ coulombs. The low value of $q^{*}_{out}$ compared to total charge is not surprising considering the fact that $q^{*}_{out}$ is due to the contribution solely from BFO nanorods protruding out of nanopores.

Let us now try to understand the charge distribution in BFO/AAO template network. Our system consists of BFO nanorods some of them protruding out of pores along with porous AAO template with pore sizes varying from 20-100nm. The combined network is schematically presented in fig. 4. Total charge will be distributed as inner charge in the pores (some of which contains BFO) and as outer charge on the BFO nanorods protruding 1 μm on the average above the surface of the template. Around half of the pores filled by BFO have protruded nanorods; others are inside the pores. The depth of the pores is 60μm– the thickness of the template. Thus, the protruded portion of BFO– solely responsible for the contribution to $q^{*}_{out}$ is 1/60 of the wt. of protruded BFO nanorod which itself is half of the total wt. 80μgms of BFO. Rest is embedded in pores. $q^{*}_{out}$ (obtained by extrapolating the plot of charge q* vs. $v^{-1/2}$ to $v^{-1/2}$ =0 and taking the intercept) is 1.50x$10^{-4}$ coulombs. On this basis, the specific capacitance of BFO nanorod structure comes out to be 450 F/gm. This large value of specific capacitance can be attributed to the nanostructure form of BFO nanorod.

The pseudocapacitative behavior of BFO stems from its redox reaction. $BiFeO_3$ is more readily reduced than oxidised, with the creation of oxygen vacancies and $Fe^{2+}$ species. The highly unfavourable energy (> 4 eV) estimated for disproportionation of $Fe^{3+}$ (to $Fe^{2+}$ and $Fe^{4+}$) suggests that tetravalent iron ions are unlikely to form in this material through this process.

## Conclusion:

We have studied capacitance of BFO nanorods developed on AAO templates by electrochemical means. Capacitance of AAO as evaluated by accumulated charge is significantly less as compared to BFO. BFO nanorods protruded from the template surface demonstrated a very high specific capacitance of 450F/gm as measured from the charge accumulated on the outer surface. The high specific capacitance is due to the particular nanostructure in the form of the rod. There is a close resemblance between the capacitance assayed by CV and charge-discharge methods. The system is quite stable with respect to repeated cycling. BFO system undergoes a redox process with O vacancies being generated giving rise to pseudocapacitative behavior with high specific capacitance. The high specific capacitance in the nanorod structure coupled with stability at long cycles brings forth its application as electrodes in batteries.

## Acknowledgements:

Authors gratefully acknowledge Dr. P.K.Mukhopadhyay and Mr. Sakti Nath Das of S.N.Bose Center for basic Sciences for SEM studies and Mr. Pulak Kumar Roy of Saha Instititute of Nuclear Physics for TEM studies.

## References:

[1] Zhang, Y.; Feng, H.; Wu, X.; Wang, L.; Zhang, A.; Xia, T.; Dong, H.; Li, X.; Linsen. Z. Int. *"Progress of electrochemical capacitor electrode materials: A review"* Jl. Hydrogen Energy 2009, **34**, 4889-4899.

[2] Zhang, X. Y.; Lai, C. W.; Zhao, X.;Wang. D. Y.; Dai, J. Y. *"Synthesis and ferroelectric properties of multiferroic BiFeO3 nanotube arrays"* Appl. Phys. Lett. 2005, **87**, 143102-143104.

[3] Lokhande, C.D.; Gujar, T.P. ; Shinde, U.R.; Mane, R.S. ; Han. S.H. "*Electrochemical supercapacitor application of pervoskite thin films"* Electrochem. Comm. 2007, 9, 1805-1807.

[4] Gao, F.; Yuan. Y.; Wing, K.F.; Chen, X.Y.; Chen, F; Liu, J.M.; Ren, Z.F; *"Preparation and photoabsorption characterization of $BiFeO_3$ nanowires"* Appl. Phys. Lett. 2006, 89 ,102506- 102508".

[5] Xie, S.H.; Li, J.Y.; Proksch, R.; Liu. Y.M.; ; Zhou, Y.C. ; Liu. Y. Y.; Pan. L.N. ; Qiao, Y. "*Nanocrystalline multiferroic $BiFeO_3$ ultrafine fibers by sol-gel based electrospinning*" Appl. Phys. Lett. 2008, 93 222904-222906.

[6] Purusothaman, K.K.; Cuba, M.; Muralidharan, "*Supercapacitor behavior of α-$MnMoO_4$ nanorods on different electrolytes"* G. Mat. Res. Bull. 2012, **47**, 3348–3351.

[7] Portet, C.; Taberna. P.L.; Simon, P. ; Laberty-Robert, C. "*Modification of Al current collector surface by sol–gel deposit for carbon–carbon supercapacitor applications"* Electrochim. Acta 2004**,** **49**, 905-912.

[8] Reddy, R. N ; Reddy, R. G., "*Electrochemical Double Layer capacitance properties of carbon in aqueous and nonaqueous electrolytes"* J. Power Source 2003, **124**, 330-337.

[9] Burke, L.D., Murphy, O.J. "*Cyclic voltammetry as a technique for determining the surface area of $RuO_2$ electrodes"* Jl. Electroanal. Chem. 1979, 96, 19-25.

[10] Ardizzone, S.; Fregonara, G.; Trasatti, S. ""Inner" and "outer" active surface of $RuO_2$ electrodes", Electrochem. Acta 1990, 35, 263-267.

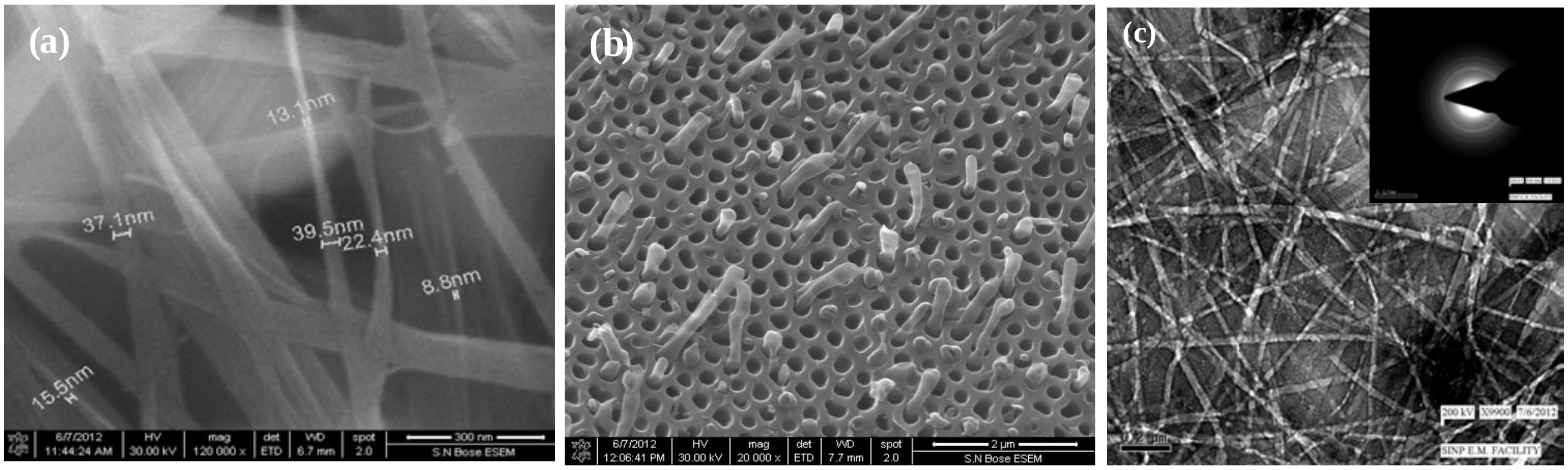

Fig. 1. (a) SEM showing bundles of nanorod, (b) SEM of nanorod protruding from pores and (c) TEM of BFO nanorods and corresponding SAED pattern shown in the inset.

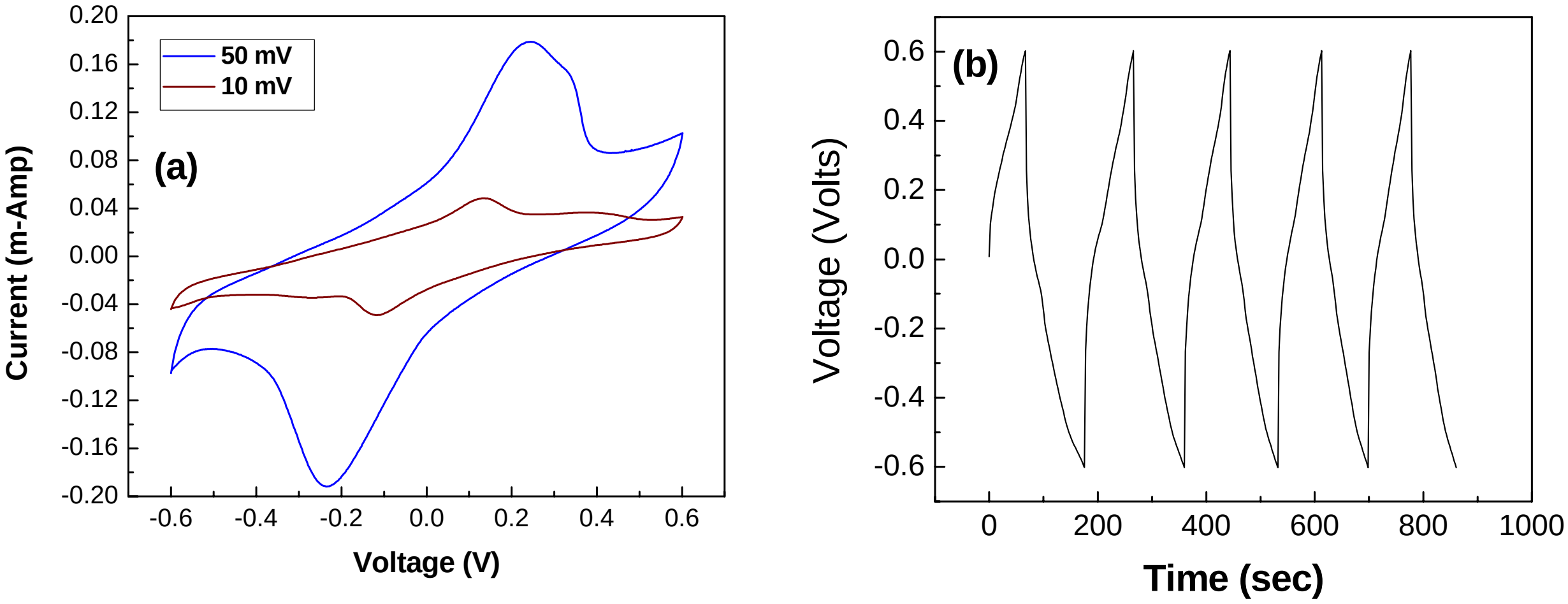

Fig. 2. (a) Cyclic voltammogram of BFO on AAO at different scan rates and (b) Charge-discharge cycle of BFO on AAO template at 30µAmp current.

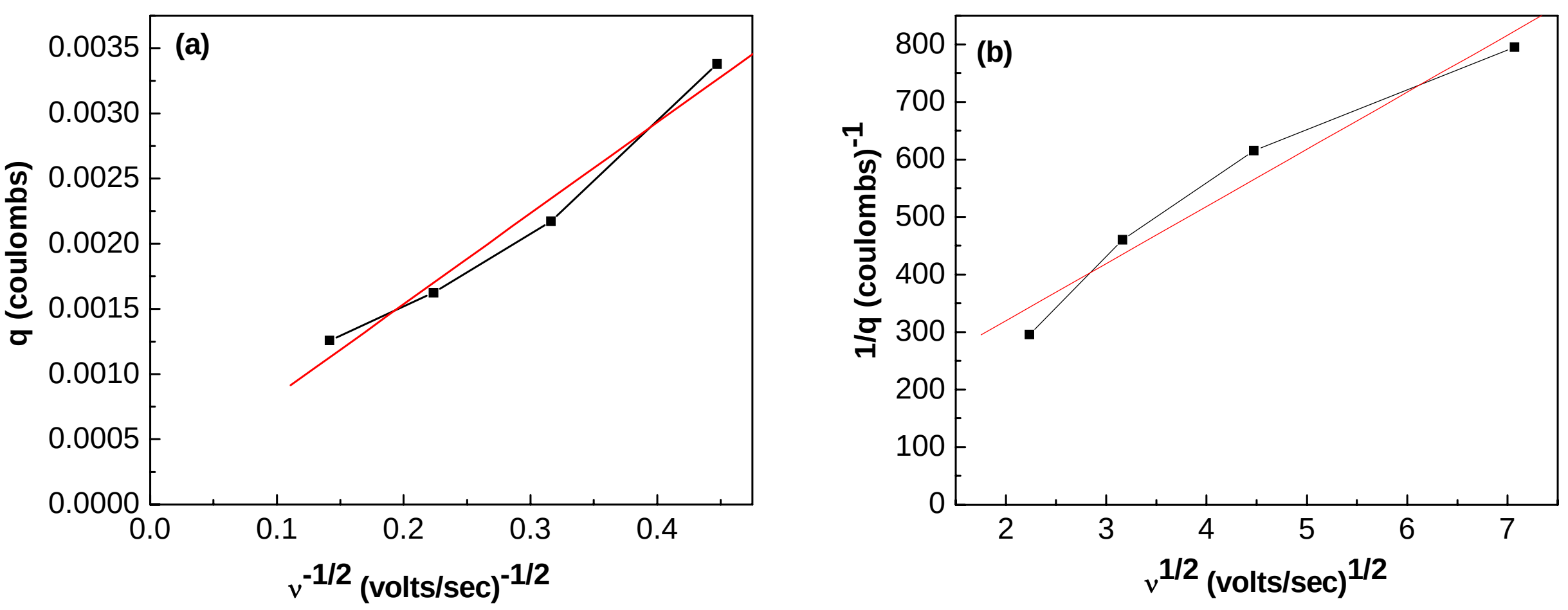

Fig. 3. (a) q* versus $v^{-1/2}$ plot to extract $q_{out}$ (b) 1/q* vs. $v^{1/2}$ to extract $q*_{total}$.

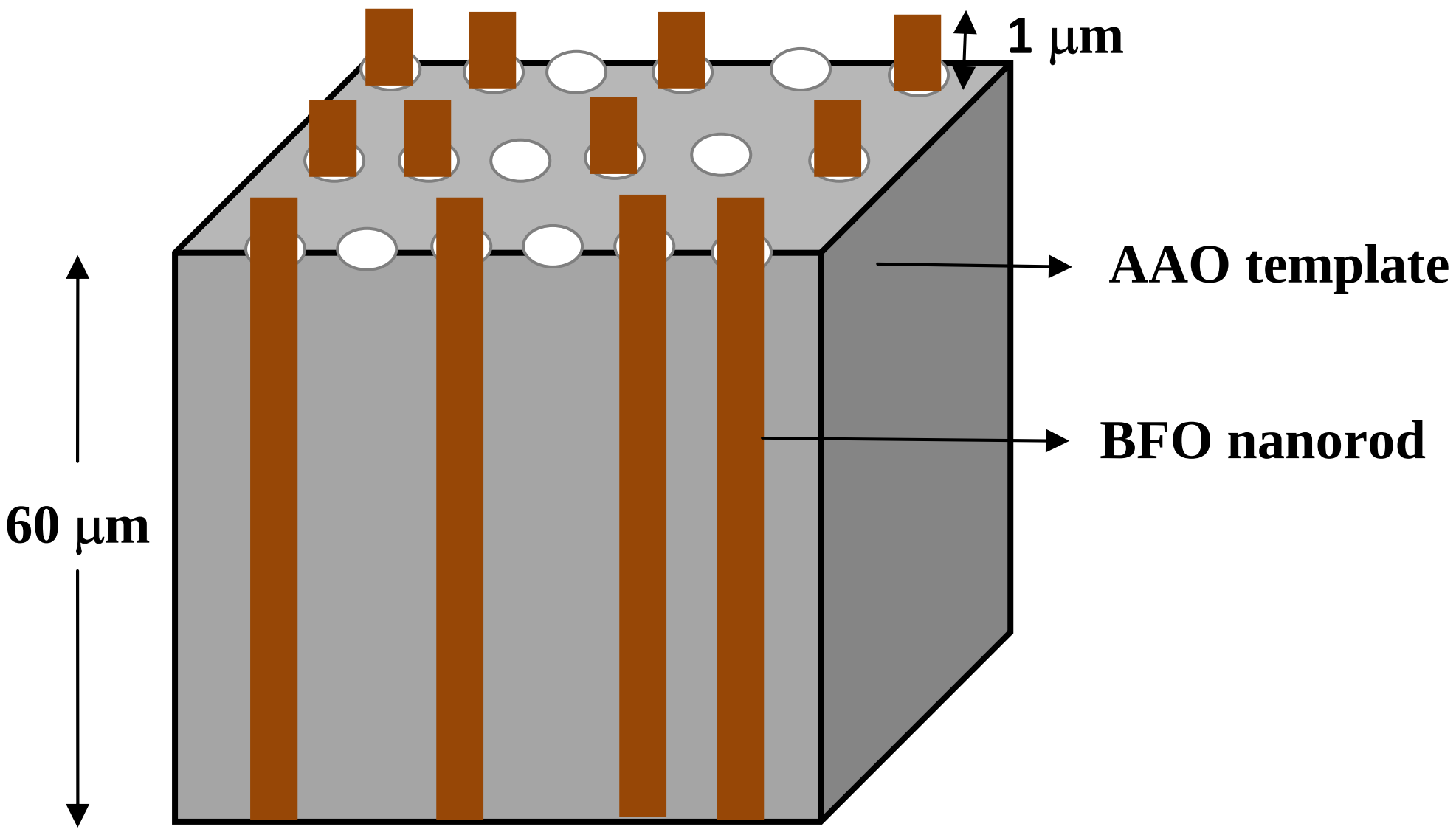

Fig. 4. Schematic diagram of BFO nanorod developed on AAO template. 1 μm protrusion from the pores is there with the depth of the pores being 60μm.